# Thermodynamics of the S$_2$-to-S$_3$ State Transition of the Oxygen-Evolving Complex of Photosystem II

Muhamed Amin[1,2,3]*, Divya Kaur[4,5], Ke R. Yang[6], Jimin Wang[7], Zainab Mohamed[8], Gary W. Brudvig[6], M.R. Gunner[4,5] and Victor Batista[6]

[1] Center for Free-Electron Laser Science, Deutsches Elektronen-Synchrotron DESY, Notkestrasse 85, 22607 Hamburg, Germany
[2] Physics Department, University College Groningen, University of Groningen, Groningen, Netherlands
[3] Centre for Theoretical Physics, The British University in Egypt, Sherouk City 11837, Cairo, Egypt
[4] Department of Physics, City College of the City University of New York, New York, NY 10031, United States
[5] Ph.D. Program in Chemistry, The Graduate Center of the City University of New York, New York, NY 10016, United States
[6] Department of Chemistry, Yale University, New Haven, Connecticut 06520-8107, United States
[7] Department of Molecular Biophysics and Biochemistry, Yale University, New Haven, CT08620-8114
[8] Zewail City of Science and Technology, Sheikh Zayed, District, 6th of October City, 12588 Giza, Egypt

Muhamed.amin@cfel.de

**Abstract**: The room temperature pump-probe X-ray free electron laser (XFEL) measurements used for serial femtosecond crystallography provide remarkable information about the structures of the catalytic (S-state) intermediates of the oxygen-evolution reaction of photosystem II. However, mixed populations of these intermediates and moderate resolution limit the interpretation of the data from current experiments. The S$_3$ XFEL structures show extra density near the OEC that may correspond to a water/hydroxide molecule. However, in the latest structure, this additional oxygen is 2.08 Å from the Oε2 of D1-E189, which is closer than the sum of the van der Waals radii of the two oxygens. Here, we use Boltzmann statistics and Monte Carlo sampling to provide a model for the S$_2$-to-S$_3$ state transition, allowing structural changes and the insertion of an additional water/hydroxide. Based on our model, water/hydroxide addition to the oxygen-evolving complex (OEC) is not thermodynamically favorable in the S$_2$ g = 2 state, but it is in the S$_2$ g = 4.1 redox isomer. Thus, formation of the S$_3$ state starts by a transition from the S$_2$ g = 2 to the S$_2$ g = 4.1 structure. Then, electrostatic interactions support protonation of D1-H190 and deprotonation of the Ca$^{2+}$-ligated water (W3) with proton loss to the lumen. The W3 hydroxide moves toward Mn4, completing the coordination shell of Mn4 and moving with its oxidation to Mn(IV) in the S$_3$ state. In addition, binding additional hydroxide to Mn1 leads to a conformational change of D1-E189 in the S$_2$ g = 4.1 and S$_3$ structures. In the S$_3$ state in the population of protonated D1-E189 increases.

## Introduction

The oxygen-evolving complex (OEC) of photosystem II (PSII) catalyzes the oxidation of water to O$_2$. With input of light, the OEC of PSII is sequentially oxidized. As the system is oxidized protons are lost. There are five OEC S-state intermediates along the catalytic cycle designated S$_0$ to S$_4$, with S$_0$ the most reduced.[1–4] The fully oxidized S$_4$ state catalyzes the oxidation of two waters to O$_2$. For understanding the reaction mechanism, it is crucial to obtain structural information about these catalytic intermediates and determine the pathway for the transitions between states.[5–10]

The OEC core contains a cluster of four Mn ions and a Ca$^{2+}$ connected through bridging oxides (*i.e.*, deprotonated water molecules).[11] In the dark-stable S$_1$ state, the Mn oxidation states are (III$_2$, IV$_2$) with the subscript indicating the number of Mn ions in a given oxidation state.[12,13] In the S$_1$-to-S$_2$ state transition, the OEC loses an electron, oxidizing an Mn(III) to Mn(IV). EPR measurements[14–17] and theoretical models[9,18] show two energetically accessible redox isomers of the S$_2$ state. When the dangler Mn4 (Figure 1) is oxidized, the Mn cluster has a total spin of S = 1/2 and produces the g = 2 multiline EPR signal ($S_{2,g=2}$). The second $S_2$ state redox isomer is formed when Mn1 (Figure 1) is oxidized and the total spin is now S = 5/2, which results in the $S_2$ state g = 4.1 EPR signal ($S_{2,g=4.1}$).[8,9,15,17,19,20] The $S_1$-to-$S_2$ state transition is accompanied by little or no proton loss to the lumen[21–23] and EXAFS measurements[24,25] show no major structural rearrangements. In contrast, the $S_2$-to-$S_3$ state transition induces loss of approximately one proton/OEC and EXAFS spectra of the $S_2$ and $S_3$ states show significant structural changes.[26,27]

Theoretical studies, recently confirmed by time-resolved XFEL structures,[28] show the insertion of an additional oxo/hydroxide in the $S_3$ state. This additional oxygen is required to complete the coordination shell of the only Mn(III) left in the $S_2$ state, which has only 5 ligands, facilitating its oxidation. Time-resolved photothermal beam deflection measurements suggested that this oxidation is preceded by a proton release from the OEC.[27,29] However, it is an open question which protonatable group (amino acids/water) are releasing a proton in this transition. Although the XFEL structures are not affected by damage due to exposure to X-ray radiation, there remains disagreement about the location of this additional oxygen reported by the different XFEL studies. This results from the inhomogeneous populations of the states of the different crystals and the time point used to probe the structures.[28,30] In addition, very high resolution is required to clearly establish the location of this oxygen near the high electron densities of the Mn ions.[31]

In the calculations reported here the position of the inserted oxygen is not fixed at the beginning of the calculation. Rather Monte Carlo sampling is used to generate a Boltzmann distribution of the possible binding sites for the oxygens of water molecules or hydroxyls, and their proton conformers in the $S_2$ g = 2, $S_2$ g = 4.1, and $S_3$ states and in the intermediate state where $Y_Z$ is oxidized and its conjugate base D1-H190 is protonated ($S_2$-$Y_Z^•$/H190$^+$).[25,32–34] $Y_Z^•$ is the electron acceptor coupled to OEC oxidation. The use of a classical, electrostatic model of the OEC enables sampling of the water/hydroxide binding sites, their protonation states and proton positions, the protonation of the amino acids and the oxidation states of the Mn centers. This method has been tested against model compounds and was used to study the proton coupled electron transfer in the Kok cycle previously.[35]

## Computational Methods

*The structural model* that includes amino acid residues and cofactors of PSII within a sphere with an ≈15 Å radius centered at the OEC cluster, two chloride ions and 85 water molecules, is optimized by QM/MM in both the g = 2 and g = 4.1 $S_2$ states.[7] The geometry optimization is carried out using the ONIOM[36] method within the Gaussian 09[37] software. The amino acid residues that are included in the 15 Å sphere are: D1 (chain A): (57)-V58-V67-(68), (81)-V82-L91-(92), (107)-N108-Y112-(113), (155)-A156-I192-(193), (289)-I290-N298-(299), (323)-A324-A344/C-terminus; CP43 (chain C):(290)-W291-(292), (305)-G306-A314-(315), (334)-T335-L337-(338), (341)-M342-(343), (350)-F351-F358-(359), (398)-A399-G402-(403), (408)-G409-E413-(414); D2 (chain D):(311)-E312-L321-(322), (347)-R348-L352/C-terminus. Only backbone atoms are considered for the capped residues highlighted in parentheses. The QM region includes the $Mn_4O_5Ca$ cluster and the amino acids ligands D1-D170, D1-E189, D1-H332, D1-E333, D1-D342, C-terminus of D1-A344, CP43-E354, D1-H337, CP43-R357, D1-D61 along with ten water molecules. The DFT/B97D[38–40] level of theory that includes dispersion correction is used for the QM region while the AMBER[41,42] force field is used for the MM region. The LanL2DZ[43,44] basis set is used for calcium and manganese metal centers while 6-31G**[45,46] is used for hydrogen, carbon, nitrogen and oxygen atoms. The capping residues, chloride ions and oxygen atoms of water molecules are frozen at the edge of the model in the MM region while all other atoms in the QM region are free to move.

*Monte Carlo (MC) sampling.* All water molecules except those that are ligands to Mn4 or $Ca^{2+}$ were removed from the QM/MM optimized structure. Then, oxygens were placed into all the cavities within 4 Å of the $Mn_4O_5Ca$ cluster on a 1.0 Å grid using the program IPECE[47] (Figure 1A). In total, 451 oxygens were added in the cavity around the OEC. Water and hydroxide proton position conformers were generated for each oxygen atom independently (7569 hydrogen conformers were added). In addition, there is one conformer per oxygen species that represents this group moving out of the protein into solvent. Thus, each of the 451 oxygen species on the grid can be either water or hydroxide or moved to solvent in the microstates that will be subjected to MC sampling.

MCCE (Multi Conformer Continuum Electrostatics)[48] was used to generate rotamers for the side chains of the amino acids. The electrostatic energies of the allowed conformational space are then calculated using Adaptive Poisson-Boltzmann Solver (APBS)[49] by solving the Poisson-Boltzmann equation. Based on the calculated energies, the Metropolis Monte Carlo method is used to generate the Grand Canonical Boltzmann distribution for all conformers as based on their electrostatic and van der Waals interactions at pH 7. The dielectric constant of the proteins is set to 4, and it is 80 for the solution around the sphere. The parameters for the OEC and ligands used here as reported previously in Amin et al.[3]

Following Grand Canonical Monte Carlo (GCMC) sampling of microstates, the structure with occupied water/hydroxyl binding sites was optimized at the DFT level using the 6-31G* basis set and the B3LYP functional. All the Mn(IV) ions were defined in the high spin state. The model included the $Mn_4O_5Ca$ cluster and the amino acids ligands (D170, E189, H332, E333, D342, A344, CP43-E354). All amino acid residues are in the D1 protein unless otherwise noted. In addition, several other residues that interact closely with the OEC were added (D61, Y161, H190, H337, CP43-R357, G171, as well as 11 crystallographic waters including water ligands of the Mn and $Ca^{2+}$ centers. The $pK_a$'s of the bridging oxides were calculated in the optimized DFT model, using MCCE as described previously.[3,18]

*EXAFS simulations*. EXAFS spectra of the $S_3$ models were computed with the FEFF program (version 6)[50] combined with the IFEFFIT code (version1.2.12).[51] We included all paths with lengths up to eight scattering legs and a Debye-Waller factor of 0.003 Å for all calculations. The energy ($E$) axis was converted into the photoelectron wave vector ($k$) space by usual transformation,

$$k = \left[\frac{2m_e}{\left(\frac{h}{2\pi}\right)^2}(E-E_0)\right]^{\frac{1}{2}},$$

where $m_e$ is the mass of the electron and $h$ is the Planck's constant, and $E_0$ = 6540.0 eV is the Fermi energy of Mn. A fractional cosine-square (Hanning) window with $dk$ = 1 was applied to the $k^3$-weighted EXAFS data. The grid of $k$ points, which are equally spaced at 0.05 Å$^{-1}$, was then used for the Fourier transformation (FT) to $R$ space. A $k$ range of 4.0 – 10.5 Å$^{-1}$ was employed for the FT of the isotropic EXAFS amplitudes. In the calculation of EXAFS for different $S_3$ state models, we allow both the simulated intensity of the EXAFS signal and the shift in the $k$-space (edge shift) to relax and find the best fit using a least squares fitting procedure as done previously.[10]

**Results and Discussions**

A)

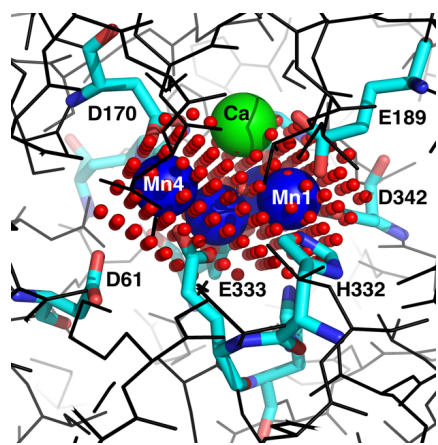

B)

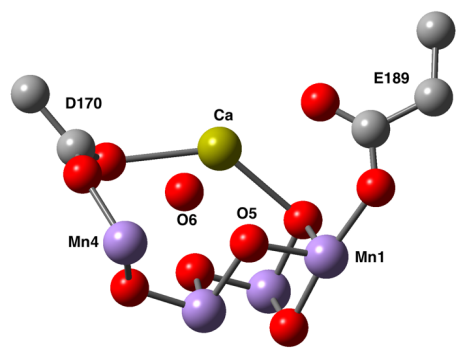

**Figure 1**. A) The cavity around the $Mn_4O_5Ca$ cluster is filled with 451 oxygens that represent water molecules or hydroxyls placed on grids with 1 Å spacing. Mn are shown in blue. B) O6 is the binding site selected by MC sampling with the classical force field.

*Water molecule-binding in the $S_2$-to-$S_3$ state transition.* The cavity around the Mn cluster in both the g = 2 and g = 4.1 $S_2$ state structures was filled with oxygen atoms of water molecules to examine the possible binding sites for an additional water/hydroxide ligand to the Mn$_4$O$_5$Ca cluster (Figure 1A). Monte Carlo sampling was used to generate a Boltzmann distribution of conformers, as defined by electrostatic and van der Waals interactions. In the S$_2$ g = 2 structure, no additional binding sites are identified nor in the S$_2$ g = 2/Y$_Z$·H190$^+$ state.

However, the Mn4-Ca$^{2+}$ distance is longer by 0.27 Å in the S$_2$ g = 4.1 structure and a hydroxide (O6 Figure 1B) that bridges Mn4 and Ca$^{2+}$ is occupied with a 25% probability, while the probability of the W3 water ligand to Ca$^{2+}$ being present is reduced from 100% to 75%. This indicates a strong coupling so that clusters will have one or the other oxygen but never both. The transfer of W3 to the hydroxide, O6 position, is coupled to the loss of a proton. The O6 oxygen is located 2.06 Å from Mn4, 2.22 Å from Ca$^{2+}$ and 1.78 Å from O5. In the S$_2$ g = 4.1/Y$_Z$·H190$^+$ state, now O6 is always present and W3 is never found in accepted microstates, which suggests a proton loss upon the formation of Y$_Z$·H190$^+$, before oxidation of the Mn cluster. This sequence of events has been observed by time-resolved photothermal beam deflection measurements.[27] The total charge of the protein upon the formation of Y$_Z$·H190$^+$ is reduced by one unit, which suggests that a proton is released to the lumen.

Cluster based DFT calculations[54] also suggested that the binding of an additional water molecule takes place at Mn1 in the g = 2 form of the S$_2$ state, a transition that is at high energy here. A similar mechanism in which W3 deprotonates and translates to complete the coordination shell of Mn4 has been proposed by Ugur et al.[52] However, they also proposed that W3 may translate toward Mn1 in the S$_2$ g = 2 structure to complete its coordination shell. This is included as a possible transition but is found to be energetically unfavorable in our simulations. Earlier computational studies generally agree that the water/hydroxide is inserted into the OEC cluster from the closed, S$_2$ state g = 4.1 structure.[8,19,32,53] The S$_2$ state g = 4.1 form is less rigid, with higher fluctuations that allow the rearrangement needed for binding an additional water molecule.[8]

The two S$_2$ structures were then advanced to the S$_3$ state to oxidize the Mn centers by changing the redox potential ($E_h$), at physiological pH where:

$$\Delta G^{Mn(III-IV)} = \left[F(E_h - E_{m,sol})\right] + \Delta\Delta G_{solvation} + \Delta\Delta G_{pairwise}$$

where $\Delta G^{Mn(III-IV)}$ is the energy required to oxidize an isolated Mn center from Mn(III) to Mn(IV), $F$ is the Faraday constant, $E_{m,sol}$ is the mid-point reduction potential of Mn(IV) in solution as obtained in Amin et al.,[35] $\Delta\Delta G_{solvation}$ is the desolvation penalty of Mn4 and $\Delta\Delta G_{pairwise}$ is the pairwise electrostatic interaction between Mn and the surrounding residues, waters, Ca$^{2+}$ and bridging oxygen atoms.

In the S$_2$ g = 4.1 structure, Mn4 is oxidized at 0.6 V and the oxidation is coupled to binding O6, while in the S$_2$ g = 2 structure Mn1 has a potential for oxidation of 1.4 V, which is higher than the potential for oxidation of P680. This suggests the advancement of the OEC to the S$_3$ state through the S$_2$ g = 2 state is energetically unfavorable, which agrees with earlier studies.[8,19,32,53,55–57] Thus, for the OEC to advance from the S$_2$ g = 2 state to the S$_3$ state, it has to transit through the S$_2$ g = 4.1 state.[58–60] Experimental studies using EPR spectroscopy[58,59] also indicates the formation of the S$_3$ state from the S$_2$ state g = 4.1 isomer at lower temperatures in both Ca-PSII and Sr-PSII. Combined multiscale *ab initio* DFT+U simulations[19] suggested that the oxidation of Y$_Z$ stabilizes the conversion of the open form (g = 2) to closed (g = 4.1) S$_2$ state isomer prior to formation of the S$_3$ state.

The analysis thus far has used a classical force field to extensively sample many oxygen and proton positions of water molecules. The S$_3$ structure with the additional hydroxide in the O6 position, obtained by MC sampling with only electrostatic and van der Waals interactions, was optimized by using DFT. The optimization adjusted the position of O6 to complete the octahedral coordination shell of Mn4 (Figure 1B), reducing the Mn4-O6 bond to 1.91 Å, while the Ca$^{2+}$-O6 and O6-O5 distances increased to 2.46 Å and 2.74 Å, respectively. Upon optimization, an additional hydrogen bond was formed between the hydroxide O6 and μ-oxo O5 with a distance of 2.30 Å.

There are two structural models of the S$_3$ state proposed in the literature, with "open" or "closed" structures of the Mn$_4$O$_5$Ca cubane that may be relevant for the catalytic reaction.[10,61–64] In the closed cubane structure, O6-H is placed 3.57 Å away from Ca$^{2+}$ and does not form a hydrogen bond with O5. In the open structure,

O6 is located at 2.97 Å from $Ca^{2+}$ forming a strong hydrogen bond with O5. The model proposed in this paper is an intermediate between the closed and open models of the $S_3$ state, with O6 forming a hydrogen bond with O5 but not as strongly as in the earlier open structure model. In addition, the O6-$Ca^{2+}$ distance is shorter in our model than the other two forms (Figure S1).

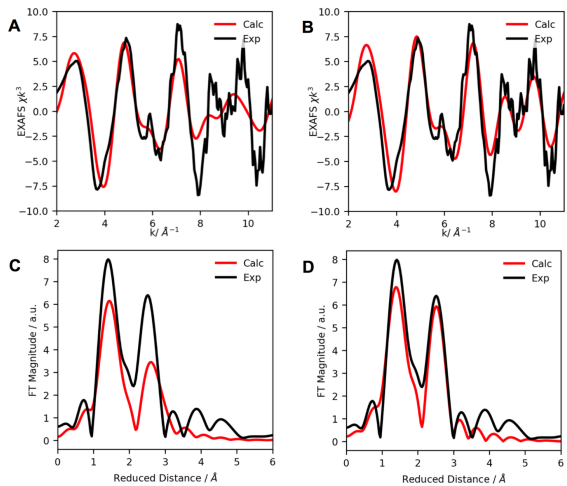

**Figure 2.** Comparison of calculated (red) and experimental (black) Mn EXAFS spectra of different $S_3$ state models in $k$-space (A and B) and reduced distance space (C and D) for the new model obtained in this study (A and C) and the previous QM/MM-optimized open form of the $S_3$ state (B and D).[10]

Figure 2 compares the EXAFS spectrum of the $S_3$ state model obtained in the current study (panels A and C) with the corresponding spectrum of the previously reported QM/MM $S_3$ model (panels B and D) and the experimental EXAFS data.[65] The comparison suggests that the $S_3$ state model obtained in the current study does not match the experimental EXAFS data as well as the QM/MM-optimized open-form of the $S_3$ state, suggesting that the structure we obtained in this study may be an intermediate structure during the S2-to-S3 state transition, with the final metastable S3 state best described as the open form of the $S_3$ state.. The largest difference in the Fourier Transformed spectrum is observed for the second peak (Figure 2 C, D), i.e. the Mn-Mn/Mn-$Ca^{2+}$ distances, which are sensitive to the different positions of the additional ligand O6 (Figure S1).

*Role of D1-E189.* While D1-E189 appears to be a ligand of Mn1 in the $S_1$ dark-adapted state, FTIR difference spectroscopy suggested that it is not a ligand of a Mn that undergoes oxidation between the $S_0$ and $S_3$ state transitions.[66] In addition, the recently published 2.07 Å resolution $S_3$ state structure by Kern et al.[28] showed a translation of E189 away from Mn1 upon the insertion of a water molecule that is only 2.09 Å away from O5 and 1.78 Å from Mn1. Thus, to examine the role of E189, we have generated more than 50 conformers of its sidechains and sampled them along with 451 water molecule-binding sites in the $S_2$ g = 2, $S_2$ g = 4.1 and $S_3$ states.

The MC sampling confirms a conformational change for E189 moving away from Mn1 that is coupled to binding another $OH^-$ (O7) i.e., O7 replacing the anionic E189 as a ligand for Mn1. These conformational changes are coupled to the oxidation of Mn1 in both the $S_2$ state g = 4.1 and the $S_3$ state structures while no conformational changes were observed for the $S_2$ state g = 2 structure. Although the XFEL structures show different conformations of E189, the conformational changes observed in our simulations are larger than the reported in the XFEL structures and suggest a complete loss of coordination of E189 from Mn1, which is replaced by a hydroxyl.

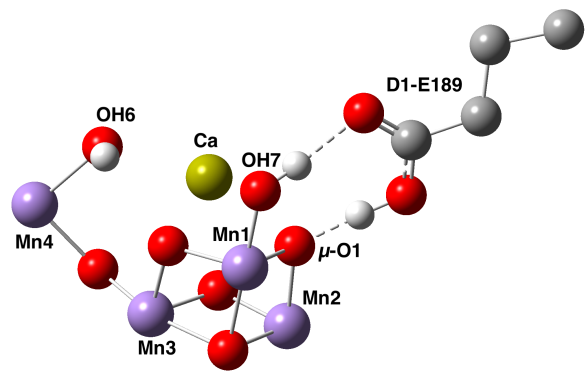

**Figure 3.** The ligand environment of Mn1 with O7 (hydroxyl anion) and protonated D1-E189. Strong hydrogen bonds (shown in dashed lines) are formed between D1-E189, OH7 and μ-O1.

The OH7 (i.e., the hydroxyl anion form of O7) has a stronger dipole moment than the carboxylate group of the amino acid and replaces the E189 ligand in the $S_2$ g = 4.1 and in the $S_3$ state due to the electrostatic attraction with Mn1. However, in the $S_2$ g = 2 state, Mn1 is in the Mn(III) state and the repulsion with the nearby oxo bridges and negatively charged ligands is stronger than the attraction to Mn1. Thus, in the $S_1$ and $S_2$ g = 2 states E189 is a Mn1 ligand and no $OH^-$ binds to Mn1. O7 binding is independent from the addition of O6, i.e. both hydroxides may bind the OEC simultaneously. However, when O7 has replaced E189 as the Mn1 ligand, the protonation state of E189 is dependent on the state of O6. Thus, E189 is ionized in the $S_2$ g = 4.1 state because O6 is not bound yet to the cluster; instead, the $Ca^{2+}$ ligand W3 is mostly occupied. In the $S_3$ state, O6 binds to the cluster and causes an approximately 50:50 mixture of protonated and deprotonated E189 due to the electrostatic repulsions.

To further assess the energetics of O7 binding, we compared the optimized DFT energies of the $S_3$ state having E189 ionized and bound to Mn1 (Mn1-E189) with the $S_3$ state having OH7 bound to Mn1 and protonated E189 (Mn1-OH7). The X, Y, Z coordinates of the two optimized structures, the DFT energies (in Hartree) and the Mn spin densities are reported in the SI (Table S1). The Mn1-OH7 structure energy is 2.4 kcal/mol lower than the Mn1-E189 structure, which indicates that the two states are close in energy, in agreement with the MC sampling calculations. The mutations of E189 (E189K, R and Q) have shown no effect on electron transfer at the donor side of photosystem II,[67] which may be explained by the existence of an isoenergetic structure (Mn1-OH7) that facilitates the oxidation of Mn1 through the Kok cycle.

*XFEL structures of the $S_3$ state*. The optimized Mn1-OH7 structure shows strong hydrogen bonds formed between O7, the carboxylate group of E189 (2.83 Å O-O distance) and the µ-oxo that bridges Mn1 and Mn2 (Figure 3). The position of O7 is very similar to the additional oxygen resolved in the latest $S_3$ structure Kern et al.[28] (Table 1). However, the Kern et al. structure also has an unusually short inter-atomic distance of 2.08 Å between the newly inserted oxygen and Oε2 of E189, which is much shorter than the sum of their van der Waals radii, and is considered to be physically impossible. Such a short distance cannot be rationalized with any known repulsive force field parameters in molecular dynamic simulation. If O7 is coordinated as a ligand of Mn1, E189 must move away from O7 through repulsive interactions with E189. It is interesting that E189 always has multiple conformations in both the Suga et al. and Kern et al. $S_3$ structures. Furthermore, Kern et al. have built two conformations for E189, and in the second of these the O7- Oε2 E189 distance is 1.60 Å. This suggests that O7 and E189 are mutually exclusive in space in this conformation. It is relatively straightforward to detect the binding of a water molecule in a location where nothing is there using $F_o$-$F_o$ isomorphous difference Fourier maps regardless of its low population. However, it is much more challenging to establish a displacement of side chains such as E189 using the same method, which may become undetectable when the population is low. This may explain the uncertainties in the sidechain positions of E189.

The $S_3$ XFEL structure by Suga et al. at 2.25 Å resolution suggested the insertion of an additional oxygen (or water substrate) that is only 1.46 Å away from O5 similar to the $S_3$ open structure.[68] However, this short distance between the two oxygens is possible only when there is H atom trapped between them and may have resulted from superposition of states with lower oxidation levels.[69] The O6 position observed in the Suga et al. $S_3$ structure is far from E189. Even so, possible multiple conformations of E189 in the $S_3$ state have been detected (for example, Figure S5 of Wang[70] et al., 2017). Therefore, the possibility of a third E189 conformation as proposed by this study does not contradict the evidence of the structure given the current uncertainty.

Table 1 shows the comparison of the $Mn_i, Ca^{2+}$-$O_j$ (i = 1,4 and j = 5,6,7) distances in the different structures. It is clear that the position of O6 is similar to the oxygen position observed in the XFEL structure by Suga et al.,[30] while the position of O7 is closer to the oxygen identified in the $S_3$ structure by Kern et al.[28] The mismatch between the XFEL measurements may have resulted from the uncertainties in the oxygen positions, due to the difficulties of accurately resolving their electron density near the heavy Mn ions, the fact that misses in a multi-flash experiment will produce a mixture of S states, the different time points used to probe the structures, and the changes induced by radiation.[71,72]

**Table 1.** The distances in the optimized $S_{2g=2}$, $S_{2g=4.1}$ and $S_3$ structures.

|  | $S_{2g=2}$ | $S_{2g=4.1}$ | $S_{3Suga}$ | $S_{3Kern}$ | $S_{3\text{-DFT}}$ |
|---|---|---|---|---|---|
| Mn1-Ca$^{2+}$ | 3.30 | 2.87 | 3.26 | 3.33 | 2.92 |
| Mn4-Ca$^{2+}$ | 3.93 | 4.20 | 4.07 | 3.90 | 3.93 |
| Mn1-O5 | 2.96 | 1.83 | 2.80 | 2.90 | 1.84 |
| Mn4-O5 | 1.82 | 2.99 | 3.51 | 2.26 | 3.35 |
| Mn1-O6 | - | - | 2.27 | - | 4.26 |
| Mn4-O6 | - | - | 2.30 | - | 1.97 |
| Mn1-O7 | - | - | - | 1.78 | 1.79 |
| Mn4-O7 | - | - | - | 4.18 | 5.74 |
| O5-O6 | - | - | 1.46 | 2.09 | 2.65 |
| O5-O7 | - | - | - | 2.10 | 2.75 |
| O6-O7 | - | - | - | - | 4.29 |

All distances are in Å. No binding sites identified for O6 or O7 in the $S_2$ g = 2 structure. The $S_{3Suga}{}^{68}$ structure is resolved at 2.5 Å (PDB ID:5GTI), while the $S_{3Kern}{}^{28}$ is resolved at 2.07 Å (PDB ID:6DHO).

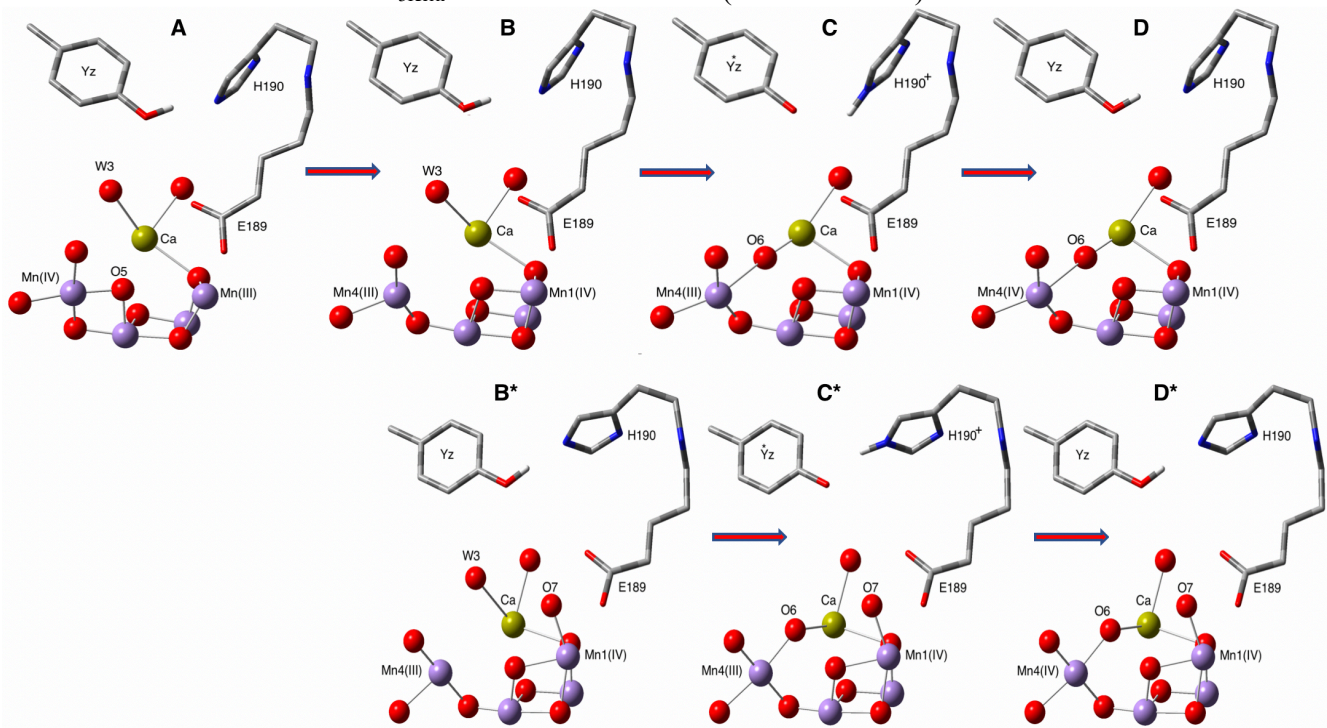

**Figure 4.** The proposed mechanism for the $S_2$-to-$S_3$ state transition. A, B, C and D are the $S_2$ g = 2, $S_2$ g = 4.1, $S_2$ g = 4.1/$Y_Z{}^\bullet$H190$^+$ and $S_3$ states, respectively. The reaction starts at A ($S_2$ g = 2) and ends at D ($S_3$). The B*, C* and D* are possible isomers of the B, C and D states, where O7 replaces D1-E189 as a ligand for Mn1.

## Conclusions

Simulations based on classical Monte Carlo sampling of many possible oxygen positions to generate the Boltzmann distribution suggest the transition from the $S_2$ to the $S_3$ state starts by a transition from the $S_2$ g = 2 state to the $S_2$ g = 4.1 state. Then, W3 on calcium deprotonates upon the formation of the $Y_Z{}^\bullet$/H190$^+$ intermediate prior to the oxidation of the Mn cluster. The deprotonation of W3 is coupled to a translation of the oxygen toward Mn4, which completes its coordination shell and facilitates the oxidation of the cluster to the $S_3$ state (Figure 4). The EXAFS spectrum of the resulting $S_3$ state structure suggests it is more consistent with an intermediate generated during the $S_2$-to-$S_3$ state transition prior to formation of the $S_3$ state in its open form.[10] In addition, the sampling of several conformers of the D1-E189 sidechain suggest that it undergoes a conformational change that is coupled to the oxidation of Mn1 and the binding another OH⁻ (O7) as suggested by the latest XFEL $S_3$ state structure (Figure 4, B*, C*, D*).

## Supporting Information

The Supporting Information includes Mn spin densities and the optimized atomic coordinates of the proposed models of the $S_3$ state.

## Acknowledgements

The authors acknowledge computational resources from NERSC (V.S.B.) and support by the U.S. Department of Energy, Office of Science, Office of Basic Energy Sciences, Division of Chemical Sciences, Geosciences, and Biosciences via Grants DESC0001423 (M.R.G. and V.S.B.), DE-FG02-05ER15646 (G.W.B.) and from European Research Council through the Consolidator Grant COMOTION (ERC-Küpper-614507).

Muhamed Amin[1,2,3]*, Divya Kaur[4,5], Ke R. Yang[6], Jimin Wang[7], Zainab Mohamed[8], Gary W. Brudvig[6], M.R. Gunner[4,5] and Victor Batista[6]

[1] Center for Free-Electron Laser Science, Deutsches Elektronen-Synchrotron DESY, Notkestrasse 85, 22607 Hamburg, Germany
[2] Physics Department, University College Groningen, University of Groningen, Groningen, Netherlands
[3] Centre for Theoretical Physics, The British University in Egypt, Sherouk City 11837, Cairo, Egypt
[4] Department of Physics, City College of the City University of New York, New York, NY 10031, United States
[5] Ph.D. Program in Chemistry, The Graduate Center of the City University of New York, New York, NY 10016, United States
[6] Department of Chemistry, Yale University, New Haven, Connecticut 06520-8107, United States
[7] Department of Molecular Biophysics and Biochemistry, Yale University, New Haven, CT08620-8114
[8] Zewail City of Science and Technology, Sheikh Zayed, District, 6th of October City, 12588 Giza, Egypt

Muhamed.amin@cfel.de

1- **Mn spin densities for optimized $S_3$ state structure.**

**Table S1.** The total energies and atomic spin densities of the Mn centers in the $S_3$ state

|  | Spin densities | |
| --- | --- | --- |
|  | Mn1-O7, E189$^0$ | Mn1-E189$^-$ |
| Mn1 | 2.942898 | 3.110379 |
| Mn2 | 3.027931 | 3.016139 |
| Mn3 | 2.939744 | 2.909465 |
| Mn4 | 3.037126 | 2.972186 |
|  | DFT (B97D) Energies (Hartree) | |
|  | -6372.99867963 | -6372.99484693 |

The difference in total energy between the 2 states is 2.4 kcal/mol. Thus, the two states are isoenergetic.

2- Figure S1. The Mn cluster in the $S_3$ state.

a) The open $S_3$ model.   b) The closed $S_3$ model.   c) The proposed $S_3$ model.

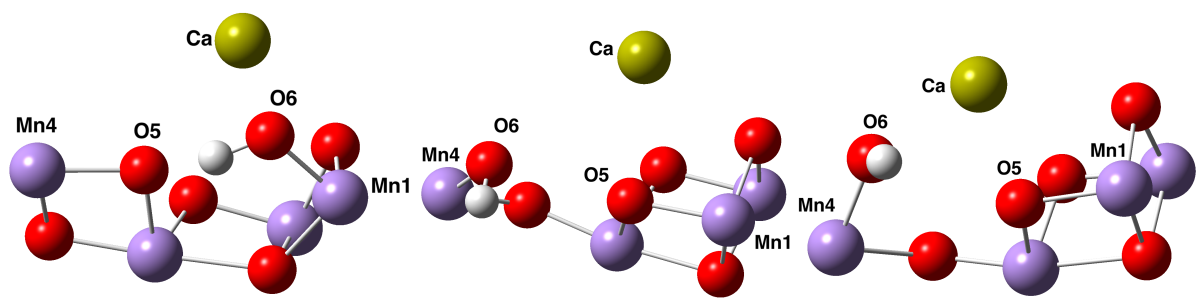

a) **The atomic coordinates of the $S_3$ state model with O7 bound to Mn1 and D1-E189 protonated:**

C       8.31218     3.59127     -1.06653

| | | | |
|---|---|---|---|
| C | 7.64049 | 3.40489 | 0.30683 |
| C | 6.13521 | 3.65811 | 0.25782 |
| O | 5.43217 | 2.89152 | 1.09338 |
| O | 5.63343 | 4.50706 | -0.48640 |
| H | 8.05668 | 4.13304 | 1.02305 |
| H | 8.08662 | 4.59153 | -1.45913 |
| H | 7.83570 | 2.40986 | 0.72654 |
| H | 9.40297 | 3.48723 | -0.97559 |
| H | 7.95388 | 2.85144 | -1.79861 |
| C | -4.42580 | 6.89976 | 0.96293 |
| C | -5.63927 | 6.02215 | 0.58770 |
| C | -5.28077 | 4.83211 | -0.28434 |
| C | -5.32517 | 3.52011 | 0.22169 |
| C | -4.90442 | 5.00220 | -1.63332 |
| C | -5.03621 | 2.40811 | -0.58269 |
| C | -4.61968 | 3.90232 | -2.45514 |
| C | -4.70803 | 2.59183 | -1.94179 |
| O | -4.49801 | 1.53268 | -2.79209 |
| H | -5.09149 | 0.73584 | -2.49237 |
| H | -3.67717 | 6.31314 | 1.51622 |
| H | -6.37617 | 6.65125 | 0.06203 |
| H | -5.61456 | 3.36299 | 1.26390 |
| H | -4.83895 | 6.01280 | -2.04224 |
| H | -5.09250 | 1.39938 | -0.17561 |
| H | -4.35750 | 4.03221 | -3.50523 |
| H | -6.13190 | 5.66160 | 1.50497 |
| C | -0.98004 | 8.14610 | 2.49347 |
| C | -0.69713 | 6.98608 | 1.53967 |
| O | -0.11515 | 5.95181 | 1.93759 |
| C | -1.71244 | 7.64952 | 3.75721 |
| H | -0.00636 | 8.57344 | 2.78017 |
| H | -1.82600 | 8.46566 | 4.48374 |
| H | -1.13864 | 6.83850 | 4.22654 |
| N | -1.12686 | 7.14126 | 0.27110 |
| C | -1.84611 | 6.70012 | -1.98557 |
| O | -2.50459 | 7.75002 | -1.90379 |
| C | -1.07205 | 6.11926 | -0.77262 |
| C | -1.70257 | 4.76655 | -0.33494 |
| C | -0.77083 | 3.56580 | -0.18377 |
| O | -1.31881 | 2.46756 | 0.07499 |
| O | 0.48854 | 3.78844 | -0.34596 |
| H | -2.46589 | 4.43935 | -1.05257 |
| H | -2.23245 | 4.88531 | 0.61548 |
| N | -1.76997 | 5.95189 | -3.11090 |
| C | -2.53233 | 6.34275 | -4.27637 |
| C | -2.42925 | 5.39121 | -5.44854 |
| O | -1.76393 | 4.36961 | -5.49841 |
| H | -3.04885 | 5.71409 | -6.32009 |
| H | -2.23696 | 7.34720 | -4.63660 |
| H | -3.60637 | 6.44435 | -4.02930 |
| H | -1.18094 | 5.10206 | -3.13476 |
| C | -6.98586 | -5.33471 | -2.33971 |
| O | -6.70918 | -6.08019 | -1.39435 |
| C | -6.10070 | -5.21774 | -3.58392 |
| C | -5.02928 | -4.11671 | -3.35990 |
| C | -4.24991 | -4.37223 | -2.06450 |
| C | -3.02784 | -3.51553 | -1.85473 |
| O | -2.44395 | -2.89230 | -2.76202 |
| O | -2.62033 | -3.52921 | -0.59358 |
| H | -3.89078 | -5.41285 | -2.01951 |
| H | -5.50962 | -3.12681 | -3.31228 |
| H | -4.89870 | -4.26368 | -1.18778 |
| H | -4.33611 | -4.10185 | -4.21251 |
| N | -8.04677 | -4.46141 | -2.31516 |
| C | -8.73169 | -4.09484 | -1.07723 |
| C | -8.88033 | -2.55556 | -0.97083 |
| C | -7.59176 | -1.81975 | -1.19497 |
| C | -7.22643 | -0.85083 | -2.11260 |
| N | -6.44612 | -2.06844 | -0.45972 |
| C | -5.45412 | -1.27069 | -0.93879 |
| N | -5.89716 | -0.51458 | -1.94230 |
| H | -8.12918 | -4.48952 | -0.24750 |
| H | -9.30379 | -2.32156 | 0.01976 |
| H | -7.84246 | -0.37289 | -2.86771 |
| H | -4.45051 | -1.24981 | -0.52850 |
| H | -9.60585 | -2.19631 | -1.71702 |
| C | 7.72881 | -2.12987 | -2.17996 |
| O | 7.72126 | -3.00585 | -1.28588 |
| C | 7.37611 | -2.49417 | -3.62497 |
| C | 5.94705 | -2.02223 | -4.02364 |
| C | 4.83549 | -2.55622 | -3.16650 |
| C | 3.74105 | -1.93992 | -2.59121 |
| N | 4.67018 | -3.90826 | -2.87640 |
| C | 3.52240 | -4.07376 | -2.16540 |
| N | 2.94782 | -2.89084 | -1.97506 |
| H | 5.77251 | -2.32443 | -5.06979 |
| H | 3.45416 | -0.89652 | -2.58688 |
| H | 3.13062 | -5.01938 | -1.81292 |
| H | 5.90935 | -0.92360 | -4.00368 |
| N | 8.02274 | -0.82762 | -1.97230 |
| C | 8.11087 | -0.23933 | -0.63061 |
| C | 6.71683 | -0.06315 | 0.01402 |
| C | 5.73436 | 0.55834 | -0.98086 |
| C | 4.28599 | 0.62270 | -0.55567 |
| O | 3.88018 | -0.19469 | 0.33989 |
| O | 3.56887 | 1.44255 | -1.21592 |
| H | 5.72057 | -0.03632 | -1.90507 |
| H | 8.74424 | -0.88313 | -0.00604 |
| H | 6.79156 | 0.56712 | 0.90967 |
| H | 6.03117 | 1.56442 | -1.28855 |
| H | 8.60386 | 0.73418 | -0.74674 |
| H | 6.34446 | -1.03613 | 0.34158 |
| C | 8.49663 | -3.71102 | 2.86210 |
| C | 7.30848 | -4.69998 | 2.82744 |
| C | 6.10886 | -4.12886 | 2.13230 |
| C | 4.78153 | -4.02480 | 2.50020 |
| N | 6.19292 | -3.54297 | 0.87476 |
| C | 4.98162 | -3.10229 | 0.50484 |
| N | 4.10795 | -3.38952 | 1.47017 |
| H | 8.82134 | -3.44933 | 1.84355 |
| H | 7.01118 | -4.97921 | 3.84806 |
| H | 4.27685 | -4.35157 | 3.40061 |
| H | 4.76776 | -2.53600 | -0.38711 |
| H | 7.61903 | -5.62930 | 2.32101 |
| H | 8.21742 | -2.78455 | 3.38419 |
| H | 9.35153 | -4.16317 | 3.38277 |
| C | -2.26391 | -6.63428 | 0.74477 |
| O | -2.44321 | -6.56495 | -0.47004 |
| C | -0.90433 | -7.03076 | 1.34760 |
| C | 0.25464 | -6.56492 | 0.44850 |
| C | 0.42231 | -5.04294 | 0.38303 |
| O | 0.06875 | -4.37679 | 1.42084 |
| O | 0.92463 | -4.57030 | -0.69063 |
| H | 0.10384 | -6.92957 | -0.57440 |
| H | 1.20326 | -6.98067 | 0.82628 |
| N | -3.25235 | -6.38560 | 1.67450 |
| C | -4.60429 | -5.96636 | 1.32658 |
| C | -4.92994 | -4.49249 | 1.63203 |
| O | -6.07689 | -4.05830 | 1.40585 |
| H | -4.75997 | -6.11855 | 0.25032 |
| N | -3.93236 | -3.74473 | 2.15345 |
| C | -4.00637 | -2.29030 | 2.32982 |
| C | -2.67565 | -1.67241 | 1.84561 |
| O | -2.66879 | -0.63493 | 1.15219 |
| C | -4.27332 | -1.90633 | 3.80020 |
| O | -1.64599 | -2.34471 | 2.27222 |
| H | -4.32771 | -0.81205 | 3.90336 |
| H | -4.81553 | -1.90392 | 1.70143 |
| H | -3.46618 | -2.28724 | 4.44110 |
| H | -3.00151 | -4.15195 | 2.18393 |
| H | -5.22948 | -2.34119 | 4.12064 |
| C | 1.67205 | 0.39540 | 7.06346 |
| C | 1.42718 | 0.12351 | 5.57124 |
| C | 2.20657 | -1.12714 | 5.08401 |
| C | 1.91660 | -1.34523 | 3.60790 |
| O | 1.03003 | -2.24375 | 3.32606 |
| O | 2.52073 | -0.58280 | 2.78631 |
| H | 1.90049 | -2.01283 | 5.65798 |
| H | 2.74125 | 0.57052 | 7.25957 |
| H | 1.73915 | 0.98916 | 4.96830 |
| H | 3.28550 | -0.95723 | 5.21516 |
| H | 1.34869 | -0.45825 | 7.67936 |
| H | 0.35263 | -0.02986 | 5.38398 |
| H | 1.11543 | 1.28498 | 7.39312 |
| C | -3.81623 | 3.42480 | 6.76441 |
| C | -3.25268 | 3.98612 | 5.44932 |
| C | -3.44560 | 3.01802 | 4.27055 |
| C | -2.84985 | 3.55277 | 2.95596 |
| N | -1.39510 | 3.74578 | 3.01003 |
| C | -0.50381 | 2.74944 | 2.87509 |
| N | -0.89203 | 1.45494 | 2.95503 |
| N | 0.80936 | 3.03884 | 2.75185 |
| H | 1.35278 | 2.34855 | 2.20788 |
| H | -3.67127 | 4.13060 | 7.59521 |
| H | -2.17747 | 4.19863 | 5.56091 |
| H | -4.52077 | 2.83475 | 4.10581 |
| H | -3.27752 | 4.54075 | 2.73916 |
| H | -1.02405 | 4.67956 | 2.78885 |
| H | -1.86137 | 1.25804 | 2.73685 |
| H | 1.02607 | 3.99327 | 2.47423 |
| H | -3.31604 | 2.47990 | 7.02978 |
| H | -3.74107 | 4.94600 | 5.20833 |
| H | -2.98915 | 2.04625 | 4.51947 |
| H | -3.10093 | 2.90824 | 2.09707 |
| H | -0.25686 | 0.74100 | 2.56907 |
| H | -4.89497 | 3.22112 | 6.67537 |
| O | 3.01985 | 3.56130 | 0.54927 |
| H | 4.42470 | 3.09736 | 0.94337 |
| H | 3.29415 | 4.35582 | 0.04690 |
| O | 2.57346 | 3.70778 | -2.03891 |
| H | 1.82012 | 3.95108 | -2.62239 |
| O | -2.35453 | -0.07691 | -2.33288 |
| H | -3.07714 | 0.54422 | -2.60452 |
| H | -2.52638 | -0.95744 | -2.71610 |
| O | -0.50313 | -2.27570 | -0.21727 |
| O | 0.24930 | -0.55314 | 1.45987 |
| O | 1.85513 | -2.39800 | 0.77569 |
| O | 1.63754 | 1.36844 | 0.74015 |
| O | 1.49278 | -0.71718 | -0.97948 |
| O | 0.02031 | 3.76108 | -3.30752 |
| H | -0.26611 | 3.57632 | -4.21681 |
| H | 0.14743 | 2.88282 | -2.84314 |
| O | 6.95536 | 1.26020 | -3.70755 |
| H | 7.38553 | 1.67166 | -4.46957 |
| H | 6.27766 | 1.93901 | -3.41483 |
| O | 5.25740 | 3.24005 | -3.02967 |
| H | 5.58142 | 3.83473 | -2.32717 |
| H | 4.30996 | 3.14432 | -2.78776 |
| Mn | 1.10915 | -2.51061 | -1.08102 |
| Mn | 0.14824 | -2.33973 | 1.53412 |
| Mn | 1.97361 | -0.36378 | 0.77653 |
| Mn | 1.99588 | 2.58241 | -0.69369 |
| Ca | -0.82566 | 0.25786 | -0.55895 |
| O | 0.74831 | 1.65260 | -1.90653 |
| H | 1.29098 | 1.03253 | -2.42160 |
| H | -6.34688 | -2.77281 | 0.29467 |
| H | 7.81815 | -0.15306 | -2.72167 |
| H | -8.15172 | -3.83407 | -3.10432 |
| H | -5.61059 | -6.18898 | -3.73318 |
| H | -6.69409 | -4.98231 | -4.48164 |
| H | -0.78737 | -6.62588 | 2.36291 |
| H | -0.87691 | -8.13003 | 1.41715 |
| H | -5.35274 | -6.58053 | 1.84785 |
| H | 7.47559 | -3.58389 | -3.72172 |
| H | 8.08795 | -2.02771 | -4.32126 |
| H | 5.31892 | -4.64367 | -3.12709 |
| H | 7.01308 | -3.38813 | 0.25751 |
| H | -4.73412 | 7.74564 | 1.59673 |
| H | -3.93736 | 7.29713 | 0.06323 |
| H | -1.55657 | 8.93923 | 1.99401 |
| H | -9.73034 | -4.55786 | -1.01998 |
| H | -3.03278 | -6.52421 | 2.65358 |
| H | -1.65668 | 7.97094 | 0.00579 |
| H | -2.71521 | 7.27196 | 3.50578 |
| H | -0.02760 | 5.93904 | -1.06598 |
| O | 0.35896 | -2.54074 | -2.70219 |
| H | -0.62101 | -2.69073 | -2.68746 |
| H | 3.08536 | -3.06936 | 1.36431 |
| H | -1.71266 | -3.03912 | -0.48479 |

b) The atomic coordinates of the S$_3$ state model with D1-E189 ionized and bound to Mn1:

| | | | |
|---|---|---|---|
| C | -3.61712 | -7.98411 | -0.49812 |
| C | -3.03985 | -7.38027 | 0.79481 |
| C | -1.74094 | -6.60961 | 0.55605 |
| O | -1.55831 | -5.58735 | 1.37737 |
| O | -0.94095 | -6.94433 | -0.33235 |
| H | -2.79961 | -8.18875 | 1.50567 |
| H | -2.86499 | -8.62343 | -0.97886 |
| H | -3.76259 | -6.72881 | 1.30331 |
| H | -4.50571 | -8.59272 | -0.27527 |
| H | -3.90402 | -7.20309 | -1.21933 |
| C | 8.01049 | -1.98665 | -1.05358 |
| C | 8.24155 | -0.88068 | -2.10544 |
| C | 6.96785 | -0.17633 | -2.54200 |
| C | 6.76459 | 1.18737 | -2.26352 |
| C | 5.96558 | -0.85402 | -3.27062 |
| C | 5.62804 | 1.87160 | -2.71888 |
| C | 4.81704 | -0.19190 | -3.72602 |
| C | 4.65350 | 1.18593 | -3.47208 |
| O | 3.56393 | 1.82780 | -4.00332 |
| H | 3.54049 | 2.81851 | -3.72999 |
| H | 7.52801 | -1.57427 | -0.15414 |
| H | 8.72994 | -1.33291 | -2.98488 |
| H | 7.52713 | 1.73541 | -1.70468 |
| H | 6.09232 | -1.91121 | -3.51065 |
| H | 5.50799 | 2.93811 | -2.52950 |
| H | 4.06453 | -0.72225 | -4.30941 |
| H | 8.94554 | -0.13402 | -1.70544 |
| C | 7.58017 | -4.20963 | 2.53845 |
| C | 6.18194 | -3.98894 | 1.96469 |
| O | 5.21646 | -3.64583 | 2.68689 |
| C | 7.54430 | -4.87446 | 3.92542 |
| H | 8.19330 | -4.79163 | 1.83327 |
| H | 8.55560 | -4.92600 | 4.35137 |
| H | 7.14167 | -5.89526 | 3.85903 |
| N | 6.05259 | -4.15689 | 0.63052 |
| C | 4.86154 | -4.50231 | -1.45871 |
| O | 5.94684 | -4.75347 | -2.01614 |
| C | 4.81247 | -3.83380 | -0.06950 |
| C | 4.62889 | -2.29281 | -0.17495 |
| C | 3.17764 | -1.84496 | -0.32823 |
| O | 2.91136 | -0.62662 | -0.31207 |
| O | 2.35972 | -2.84402 | -0.41680 |
| H | 5.20773 | -1.86682 | -1.00584 |
| H | 5.00136 | -1.82813 | 0.74783 |
| N | 3.67314 | -4.78738 | -2.03736 |
| C | 3.66420 | -5.45577 | -3.33349 |
| C | 4.52134 | -4.70330 | -4.36428 |
| O | 4.29340 | -3.54694 | -4.69952 |
| H | 5.35621 | -5.27283 | -4.82545 |
| H | 2.61905 | -5.46639 | -3.67390 |
| H | 4.03539 | -6.48787 | -3.23963 |
| H | 2.77867 | -4.48531 | -1.64213 |
| C | -1.01506 | 7.58419 | -1.95121 |
| O | -1.29141 | 8.44567 | -1.10585 |
| C | -2.01755 | 6.49977 | -2.38036 |
| C | -1.40054 | 5.10668 | -2.60997 |
| C | -2.47567 | 4.01653 | -2.81066 |
| C | -1.80924 | 2.69526 | -3.16216 |
| O | -1.03843 | 2.61585 | -4.13220 |
| O | -2.06287 | 1.62285 | -2.42712 |
| H | -3.12113 | 4.29197 | -3.66164 |
| H | -0.77950 | 4.83841 | -1.74593 |
| H | -3.09798 | 3.91020 | -1.91606 |
| H | -0.76231 | 5.10159 | -3.50726 |
| N | 0.19284 | 7.54807 | -2.59012 |
| C | 1.31105 | 8.40628 | -2.19844 |
| C | 2.58761 | 8.00084 | -2.97030 |
| C | 2.90068 | 6.53653 | -2.84467 |
| C | 3.08366 | 5.54123 | -3.78887 |
| N | 2.99580 | 5.89175 | -1.62114 |
| C | 3.22644 | 4.57310 | -1.85731 |
| N | 3.29266 | 4.32595 | -3.16418 |
| H | 1.48859 | 8.31415 | -1.11623 |
| H | 3.41957 | 8.61674 | -2.59273 |
| H | 3.08881 | 5.63528 | -4.87021 |
| H | 3.31228 | 3.83282 | -1.06920 |
| H | 2.47186 | 8.23184 | -4.03956 |
| C | -7.05607 | -3.40564 | -1.28931 |
| O | -7.52258 | -2.72175 | -0.34446 |
| C | -7.20217 | -2.93288 | -2.73607 |
| C | -5.89040 | -2.27583 | -3.26041 |
| C | -5.36781 | -1.16098 | -2.40332 |
| C | -4.10881 | -0.90347 | -1.90491 |
| N | -6.14347 | -0.08181 | -1.98494 |
| C | -5.36289 | 0.77407 | -1.27394 |
| N | -4.12665 | 0.28986 | -1.20230 |
| H | -6.07882 | -1.90371 | -4.28106 |
| H | -3.18684 | -1.45533 | -2.03671 |
| H | -5.69987 | 1.69813 | -0.82326 |
| H | -5.11039 | -3.04599 | -3.34101 |
| N | -6.38560 | -4.56203 | -1.12020 |
| C | -5.93809 | -5.02451 | 0.20025 |
| C | -4.75162 | -4.18508 | 0.72721 |
| C | -3.69810 | -4.01319 | -0.36803 |
| C | -2.49291 | -3.15142 | -0.06278 |
| O | -2.57367 | -2.28097 | 0.87628 |
| O | -1.51776 | -3.32392 | -0.85624 |
| H | -4.15603 | -3.54109 | -1.24722 |
| H | -6.78573 | -4.96977 | 0.89551 |
| H | -4.30911 | -4.66786 | 1.60885 |
| H | -3.32466 | -4.97901 | -0.71897 |
| H | -5.64124 | -6.07392 | 0.08145 |
| H | -5.13475 | -3.21309 | 1.04937 |
| C | -9.35234 | -0.85467 | 2.14347 |
| C | -8.62787 | 0.29688 | 1.40874 |
| C | -7.13926 | 0.24958 | 1.57969 |
| C | -6.22228 | 1.17436 | 2.03368 |
| N | -6.40642 | -0.88941 | 1.26111 |
| C | -5.10572 | -0.66651 | 1.49354 |
| N | -4.97589 | 0.57220 | 1.98085 |
| H | -9.03986 | -1.82639 | 1.73667 |
| H | -8.98549 | 1.27323 | 1.76534 |
| H | -6.36262 | 2.18411 | 2.39680 |
| H | -4.27117 | -1.33172 | 1.28735 |
| H | -8.86663 | 0.23706 | 0.33311 |
| H | -9.12990 | -0.83177 | 3.21972 |
| H | -10.43853 | -0.76086 | 2.00878 |
| C | -1.70133 | 6.27740 | 1.50928 |
| O | -1.23506 | 5.73863 | 0.50670 |
| C | -3.13115 | 6.03050 | 2.01258 |
| C | -3.88752 | 5.03792 | 1.11975 |
| C | -3.23370 | 3.65736 | 1.03685 |
| O | -2.40420 | 3.35035 | 1.95901 |
| O | -3.59794 | 2.90773 | 0.06296 |
| H | -3.96169 | 5.41827 | 0.09255 |
| H | -4.91690 | 4.91040 | 1.49253 |
| N | -0.94926 | 7.13000 | 2.29186 |
| C | 0.27202 | 7.72955 | 1.74550 |
| C | 1.44979 | 6.76467 | 1.54905 |
| O | 2.39622 | 7.11600 | 0.82049 |
| H | 0.06749 | 8.17517 | 0.76042 |
| N | 1.41437 | 5.59269 | 2.22986 |
| C | 2.32035 | 4.47727 | 1.95545 |
| C | 1.50592 | 3.26850 | 1.43603 |
| O | 1.92458 | 2.58609 | 0.48240 |
| C | 3.10872 | 4.05962 | 3.21447 |
| O | 0.43735 | 3.06268 | 2.15276 |
| H | 3.79096 | 3.22998 | 2.97075 |
| H | 3.01465 | 4.79881 | 1.17210 |
| H | 2.41625 | 3.73689 | 4.00513 |
| H | 0.54699 | 5.35375 | 2.70160 |
| H | 3.70531 | 4.90592 | 3.58018 |
| C | 0.38681 | -1.41159 | 7.19023 |
| C | 0.25316 | -1.00270 | 5.71549 |
| C | -1.12291 | -0.34788 | 5.42949 |
| C | -1.24785 | -0.00697 | 3.95126 |
| O | -1.19479 | 1.24026 | 3.63216 |
| O | -1.34904 | -0.99530 | 3.15069 |
| H | -1.24915 | 0.56374 | 6.02941 |
| H | -0.39378 | -2.13720 | 7.46698 |
| H | 0.36421 | -1.88074 | 5.06289 |
| H | -1.91993 | -1.06222 | 5.68750 |
| H | 0.28815 | -0.53900 | 7.85498 |
| H | 1.05060 | -0.29539 | 5.43929 |
| H | 1.36581 | -1.87428 | 7.38088 |
| C | 3.91338 | -0.24127 | 7.69247 |
| C | 4.41184 | -0.69924 | 6.31248 |
| C | 4.42122 | 0.45359 | 5.29519 |
| C | 4.95646 | 0.03961 | 3.90916 |
| N | 4.25605 | -1.10305 | 3.31606 |
| C | 2.96278 | -1.09549 | 2.93761 |
| N | 2.24578 | 0.04767 | 2.96738 |
| N  |  2.36662 | -2.24731 |  2.59294 |
| H  |  1.48572 | -2.22196 |  2.04873 |
| H  |  3.89496 | -1.07624 |  8.40785 |
| H  |  3.76818 | -1.50773 |  5.93097 |
| H  |  5.05317 |  1.27845 |  5.66362 |
| H  |  6.00819 | -0.26584 |  3.99118 |
| H  |  4.73716 | -2.00883 |  3.26690 |
| H  |  2.76474 |  0.91470 |  2.98403 |
| H  |  2.95123 | -3.07333 |  2.51050 |
| H  |  2.89314 |  0.16570 |  7.62117 |
| H  |  5.42883 | -1.11934 |  6.39793 |
| H  |  3.39987 |  0.85357 |  5.19774 |
| H  |  4.92861 |  0.89476 |  3.21001 |
| H  |  1.34864 |  0.09891 |  2.44309 |
| H  |  4.56430 |  0.54574 |  8.10399 |
| O  |  0.55005 | -4.57962 |  0.50720 |
| H  | -0.65487 | -5.08190 |  1.08291 |
| H  |  0.70926 | -5.34561 | -0.07929 |
| O  |  0.72585 | -4.20909 | -2.09679 |
| H  |  0.95767 | -3.66463 | -2.89565 |
| O  |  1.10663 |  1.23980 | -3.14925 |
| H  |  1.98961 |  1.39168 | -3.57879 |
| H  |  0.41843 |  1.73415 | -3.65170 |
| O  | -0.96795 |  2.25648 | -0.06600 |
| O  |  0.07720 |  0.47286 |  1.42256 |
| O  | -2.46417 |  0.73089 |  1.29036 |
| O  |  0.13370 | -1.99914 |  0.92060 |
| O  | -1.42066 | -0.30148 | -0.65456 |
| O  |  1.57807 | -2.82558 | -4.35183 |
| H  |  2.55437 | -2.93197 | -4.33947 |
| H  |  1.30837 | -3.21250 | -5.19707 |
| O  | -4.39571 | -5.35978 | -3.06061 |
| H  | -4.52111 | -5.96118 | -3.80734 |
| H  | -3.40634 | -5.39521 | -2.88695 |
| O  | -1.75072 | -5.68885 | -2.73652 |
| H  | -1.53439 | -6.34213 | -2.04153 |
| H  | -1.10271 | -4.97691 | -2.54529 |
| Mn | -2.39500 |  1.25605 | -0.54371 |
| Mn | -1.01528 |  1.87246 |  1.72775 |
| Mn | -1.18024 | -0.86960 |  1.09644 |
| Mn |  0.45138 | -3.04563 | -0.66162 |
| Ca |  0.93884 |  0.61737 | -0.86826 |
| O  |  0.47756 | -1.58515 | -1.77157 |
| H  | -0.25666 | -1.53967 | -2.44018 |
| H  |  2.85757 |  6.32996 | -0.69778 |
| H  | -5.85114 | -4.93909 | -1.91689 |
| H  |  0.41594 |  6.74667 | -3.16952 |
| H  | -2.76461 |  6.45539 | -1.57847 |
| H  | -2.53218 |  6.83725 | -3.29672 |
| H | -3.08278 | 5.65786 | 3.04676 |
| H | -3.66939 | 6.99166 | 2.03241 |
| H | 0.59877 | 8.52610 | 2.42792 |
| H | -8.03952 | -2.22220 | -2.76481 |
| H | -7.44834 | -3.77527 | -3.39803 |
| H | -7.13786 | 0.01947 | -2.14140 |
| H | -6.79696 | -1.72715 | 0.76675 |
| H | 8.96527 | -2.44898 | -0.75560 |
| H | 7.36227 | -2.77361 | -1.46286 |
| H | 8.04971 | -3.21320 | 2.61284 |
| H | 1.06354 | 9.45827 | -2.40424 |
| H | -1.42886 | 7.64680 | 3.01967 |
| H | 6.84763 | -4.41624 | 0.05297 |
| H | 6.89887 | -4.29845 | 4.60161 |
| H | 3.97885 | -4.24708 | 0.50983 |
| O | -1.22501 | -0.80859 | -3.73300 |
| H | -1.53551 | 0.01640 | -3.30095 |
| H | -0.41722 | -0.52436 | -4.19266 |
| H | -4.01345 | 0.97248 | 2.01379 |